# The effects of V doping on the intrinsic properties of $SmFe_{10}Co_2$ alloys: a theoretical investigation

Diana Benea[a], Viorel Pop[a] and Jan Minár[b]

[a]*Faculty of Physics, Babeş-Bolyai University Cluj-Napoca, Kogãlniceanu str 1,*

*400084 Cluj-Napoca, Romania*

[b]*New Technologies–Research Center, University of West Bohemia, 30100 Pilsen, Czech Republic*

## Abstract

The present study focuses on the intrinsic properties of the $SmFe_{10}Co_{2-x}V_x$ (x = 0-2) alloys, which includes the $SmFe_{10}Co_2$ alloy, one of the most promising permanent magnets with the $ThMn_{12}$ type of structure due to its large saturation magnetization ($\mu_0 M_s$ = 1.78 T), high Curie temperature ($T_c$ = 859 K), and anisotropy field ($\mu_0 H_a$ = 12 T) experimentally obtained. Unfortunately, its low coercivity (<0.4 T) hinders its use in permanent magnet applications. The effect of V-doping on magnetization, magnetocrystalline anisotropy energy, and Curie temperature is investigated by electronic band structure calculations. The spin-polarized fully relativistic Korringa-Kohn-Rostoker (SPR-KKR) band structure method, which employs the coherent potential approximation (CPA) to deal with substitutional disorder, has been used. The Hubbard-U correction to local spin density approximation (LSDA +U) was used to account for the large correlation effects due to the *4f* electronic states of Sm. The computed magnetic moments and magnetocrystalline anisotropy energies were compared with existing experimental data to validate the theoretical approach's reliability. The exchange-coupling parameters from the Heisenberg model were used for obtaining the mean-field estimated Curie temperature. The magnetic anisotropy energy was separated into contributions from transition metals and Sm, and its relationships with the local environment, interatomic distances, and valence electron delocalization were analyzed. The suitability of the hypothetical $SmFe_{10}CoV$ alloy for permanent magnet manufacture was assessed using the calculated anisotropy field, magnetic hardness, and intrinsic magnetic properties.

## 1. Introduction

The $SmFe_{12}$-based compounds with $ThMn_{12}$ structure type (space group I4/mmm) are considered excellent candidates for permanent magnet (PM) applications due to their intrinsic magnetic properties combined with a low ratio of rare earth and high content of low-cost iron (> 75%) [1, 2]. The partial substitution of Fe with other transition metals M (Ti, V, Cr, Mo, W) or main group elements (Al, Si, Ga) in $SmFe_{12-x}M_x$ is necessary to stabilize the phase, as binary Sm - Fe intermetallic phase is unstable and only thin films can be produced, which cannot serve as permanent magnets [1]. Further improvement of stability as well as the adjustment of the intrinsic and extrinsic parameters can be obtained by Sm substitution with Zr, Gd, and Y [2-5].

One of the most promising $ThMn_{12}$ structure type systems for PM applications is the $SmFe_{9.6}Co_{2.4}$ alloy, produced as a thin film with ~ 600 nm thickness on V substrate by Hirayama et al. [6]. The film has a Curie temperature $T_c$ = 859 K, saturation magnetization $\mu_0M_s$ = 1.78 T, and the anisotropy field $\mu_0H_a$ = 12 T [6], superior to those of $Nd_2Fe_{14}B$. Unfortunately, this compound has not yet been synthesized into a bulk alloy suitable for PM manufacturing. Moreover, later studies show low values of coercivity for the epitaxial films of these alloys ($\mu_0H_c$ only 0.1-0.4 T) [7]. Special techniques to optimize the microstructure by infiltration of non-magnetic alloys (Cu and Ga) performed by Ogawa et al. [8] enhanced $\mu_0H_c$ at ~ 0.8 T. Another procedure to reach high $\mu_0H_c$ by co-sputtering of non-magnetic boron resulted in a substantial increase of coercivity of ~1.2 T in 100 nm thick layers [9].

On the other hand, recent studies dedicated to $SmFe_{12-x}V_x$ (x = 0.5-2) alloys [10] show that bulk samples of these alloys can be obtained for x ≥ 1 by classical route consisting of arc-melting of pure metals, followed by annealing. The values of intrinsic magnetic properties experimentally obtained ($\mu_0M_s$ = 1.12 T, $\mu_0H_a$ =11 T and $T_c$ = 634 K for $SmFe_{11}V$ and, respectively, $\mu_0M_s$ = 0.81 T, $\mu_0H_a$ = 9.8 T and $T_c$ = 595 K for $SmFe_{10}V_2$) [10] are lower than the corresponding values for $SmFe_{9.6}Co_{2.4}$ alloy, but still in view for further developments in permanent magnets field. In addition, one has to stress that later studies on $SmFe_{12-x}V_x$ (x = 0 – 1.9) alloys produced either in bulk by induction melting or as thin films, show superior values of $\mu_0M_s$, $\mu_0H_a$, and $T_c$ in $SmFe_{11}V$ compared with $SmFe_{11}Ti$ [11]. As a consequence, V is more beneficial compared with Ti as phase stabilizing element M in the $Sm(Fe,M)_{12}$ compounds, due to better intrinsic magnetic properties at similar atomic percentages. Furthermore, earlier research demonstrate that $SmFe_{12-x}V_x$ alloys are susceptible to the development of viable coercivities, a crucial extrinsic property that is strongly demanded for the practical realization of permanent magnets. As shown in earlier studies, so-called bulk-hardening is employed for $Sm(Fe,M)_{12}$ alloys, via a simple procedure consisting of recrystallization of regularly cast ingots. For the $SmFe_{10}V_2$ alloy, previous research reported a coercivity value of 0.37T [12, 13]. According to other research, heat treatment applied after mechanical alloying can raise the coercivity of the Sm-Fe-V 1:12 phase to 1.17 T [14], which is one of the highest values ever achieved for 1:12 alloys. More recent studies for Sm-Fe-V 1:12 alloy obtained by hot compaction of mechanically milled powders found a maximum coercivity of 1.06 T [15, 16].

A functional permanent magnet is likely to be obtained between the $SmFe_{10}Co_{2-x}V_x$ alloys given (i) the improve of the intrinsic magnetic properties due to Co addition, (ii) the phase stability (for x >1) and relatively large coercivity due to V addition. On the other hand, the V-doping is likely to alter the intrinsic properties of the $SmFe_{10}Co_{2-x}V_x$ alloys, whereas increased Co concentration is detrimental to phase stability. As consequence, the right balance between these two contributions has to be achieved. Our theoretical investigations are focused on the evolution of the intrinsic magnetic properties by V doping in the $SmFe_{10}Co_{2-x}V_x$ (x = 0-2) alloys. The total magnetic moments, magnetic anisotropy energies, and Curie temperatures for the $SmFe_{10}V_2$ and $SmFe_{10}Co_2$ alloys are calculated, allowing for a direct comparison between their performances as permanent magnets, as well as a comparison with previous theoretical and experimental results. The dependence of the anisotropy field $H_a$ on V-doping is also investigated, as an increase in $H_a$ would allow for a higher upper limit for coercivity, according to Kronmüller's empiric equation, which is a necessary prerequisite for a functional permanent magnet.

## 2. Calculation details

Electronic band structure calculations were performed using the spin-polarized fully relativistic Korringa-Kohn-Rostoker (SPR-KKR) band structure method, which is based on the KKR-Green's function formalism that makes use of multiple scattering theory [17]. The fully relativistic approach has been employed, i.e., all relativistic effects have been taken into account, including the spin-orbit coupling. The atomic sphere approximation (ASA) was used, assuming overlapping atomic spheres inside which the electronic charge is spherically symmetric [18]. The angular momentum expansion of the basis functions was taken up to $l = 3$ for Sm and $l = 2$ for Fe, Co, and V. The exchange and correlation effects have been accounted for by employing the local spin density approximation (LSDA) with the parametrization of Vosko, Wilk, and Nusair [19]. The k-space integration was performed using the special points method [20]. The substitutional disorder in the system has been treated within the coherent potential approximation (CPA) theory [21].

Additionally, electron correlations that are not fully regarded in the LSDA have been systematically accounted for beyond the LSDA. The LSDA+U approach [17] has been used to account for on-site Coulomb interactions caused by the localized *4f* electrons of Sm. The double-counting was treated by the so-called atomic limit expression of Czyżyk and Sawatsky [22]. The parametrization of the Hubbard U and Hund exchange J parameters used values U = 6.0 eV and J = 0.9 eV, sufficient to split the 4f orbital bands into lower and upper Hubbard bands. Similar values of U and J have been used in previous LSDA+U theoretical calculations for Sm in 1:12 alloys [23, 24].

Magnetic anisotropy was studied by computing the magnetic torque acting on the magnetic moment $\overrightarrow{m_i}$ of the atomic site i, orientated along the magnetization direction $\vec{M}$ [25, 26]. The component of the magnetic torque with respect to the axis $\hat{u}$ is defined by $T_{\hat{u}}(\theta,\varphi) = -\partial E(\overrightarrow{M(\theta,\varphi)})/\partial\theta$, where $\theta$ and $\varphi$ are the polar angles. A special geometry can be used to relate the energy difference between the in-plane and out-of-plane magnetization directions and the magnetic torque. For a uniaxial anisotropy, by setting the polar angles to θ = π/4 and φ = 0, the calculated magnetic torque is the energy difference between in-plane and out-of-plane magnetization by the formula $T_{\hat{u}}(\pi/4,0) = E_{[100]} - E_{[001]}$ [26]. On the other hand, the magnetic anisotropy energy can be approximated in first order using the formula $E_a = K_1 sin^2\theta + K_2 sin^4\theta$, where $K_1$ and $K_2$ are the anisotropic constants and $\theta$ is the polar angle between the magnetization direction and the easy axis. According to the $E_a$ expression, the magnetic anisotropy energy defined as the energy difference between in-plane and out-of-plane magnetization is given by $K_u \approx K_1 + K_2$. A dense k-mesh of *25×25×25* k-points has been used for magnetic torque calculations.

The magnetic behaviour of solids is further investigated based on a microscopic model making use of magnetic interactions. In particular, the presented approach is based on the classical Heisenberg Hamiltonian described by the expression:

$$H_{ex} = -\sum_{ij} J_{ij} \hat{e}_i \cdot \hat{e}_j \qquad (1)$$

where the summation is performed on all lattice sites $i$ and $j$ and $\hat{e}_i / \hat{e}_j$ are the unit vectors of magnetic moments on sites *i* and *j,* respectively. The $J_{ij}$ exchange coupling parameters between spin magnetic moments were determined using the Liechtenstein formulation [27], which is based on the magnetic force theorem. The calculation of total energy difference $\Delta E_{ij}$ due to infinitesimal change in angle between magnetic moments of (*i, j*) spin pair is performed in scalar-relativistic mode and $\Delta E_{ij}$ dependence on the angle between $\hat{e}_i$ and $\hat{e}_j$ is derived. Following this derivation, a one-to-one mapping between the exchange-coupling energy $\Delta E_{ij}$ and the Heisenberg Hamiltonian is realized, allowing to obtain the $J_{ij}$ exchange coupling parameters [18,24]. An essential advantage of the SPR-KKR technique is the use of Liechtenstein's formula, which makes it possible to calculate the ensemble-averaged coupling constants $J_{ij}$ directly using the CPA formalism.

The Curie temperatures have been calculated within the mean-field approach [27, 28], using the expression [29]:

$$T_c^{rough-MFA} = \frac{2}{3k_B} \sum_i J_{0i} \qquad (2)$$

In this expression, the $J_{0i}$ exchange-coupling parameters are obtained by $J_{ij}$ summation over all coordination shells up to 25 Å around lattice site *i*. The Curie temperature is obtained by summation over all lattice sites.

## 3. Results and discussions

The $Sm(Fe,M)_{12}$ compounds (M = Co, V) crystallize in the $ThMn_{12}$ structure (space group *I4/mmm*) having Fe/M atoms on 3 inequivalent crystal sites (*8i*, *8j* and *8f*) and the Sm atoms on *2a* sites (Fig. 1). The V atoms occupy preferentially the *8i* sites [1,11], whilst the Co atoms occupy the 8f sites, according to the theoretical calculations of Odkhuu [23]. The experimental lattice constants for the investigated $SmFe_{10}Co_{2-x}V_x$ alloys are summarized in Table 1. The lattice constant $a_{lat}$ has the lowest values for $SmFe_{10}Co_2$, which increases by V for Co substitution [6,10]. The values of the lattice constants agree with the behaviour of the metallic radii, which decrease from V to Fe and Co. The $c_{lat}$ dependency of lattice constants on M metallic radii is not evidenced by the experimental data in the investigated doping range, since the same $c_{lat}$ value was reported for $SmFe_{10}V_2$ and $SmFe_{10}Co_2$ alloys [6, 10]. The internal coordinates $x_{8i} = 0.361$ and $x_{8j} = 0.275$ for $SmFe_{10}V_2$ [30] have been used for all investigated alloys. Accounting for the stability of $Sm(Fe,V)_{12}$ alloys for V content required to stabilize the phase [13], the $SmFe_{10}CoV$ alloy is likely to be obtained. The lattice constants for $SmFe_{10}CoV$ were approximated accounting for a linear dependence with V/Co ratio between $SmFe_{10}Co_2$ and $SmFe_{10}V_2$ lattice constants. The lattice constants and the muffin-tin radii (RMT) for *8i*, *8j* and *8f* sites of the investigated alloys are displayed in Table1.

*Table 1. Lattice constants and the muffin-tin radii (RMT) for 8i, 8j and 8f sites of the* $SmFe_{10}Co_{2-x}V_x$ *(x = 0-2) alloys.*

| | $a_{lat}$ (Å) | $c_{lat}$ (Å) | *RMT 8i (Å)* | *RMT 8j (Å)* | *RMT 8f (Å)* |
|---|---|---|---|---|---|
| $SmFe_{10}V_2$ | 8.53 [10] | 4.77 [10] | 1.196 | 1.225 | 1.191 |
| $SmFe_{10}CoV$ | 8.47 | 4.77 | 1.189 | 1.221 | 1.192 |
| $SmFe_{10}Co_2$ | 8.42 [6] | 4.77 [6] | 1.180 | 1.217 | 1.191 |

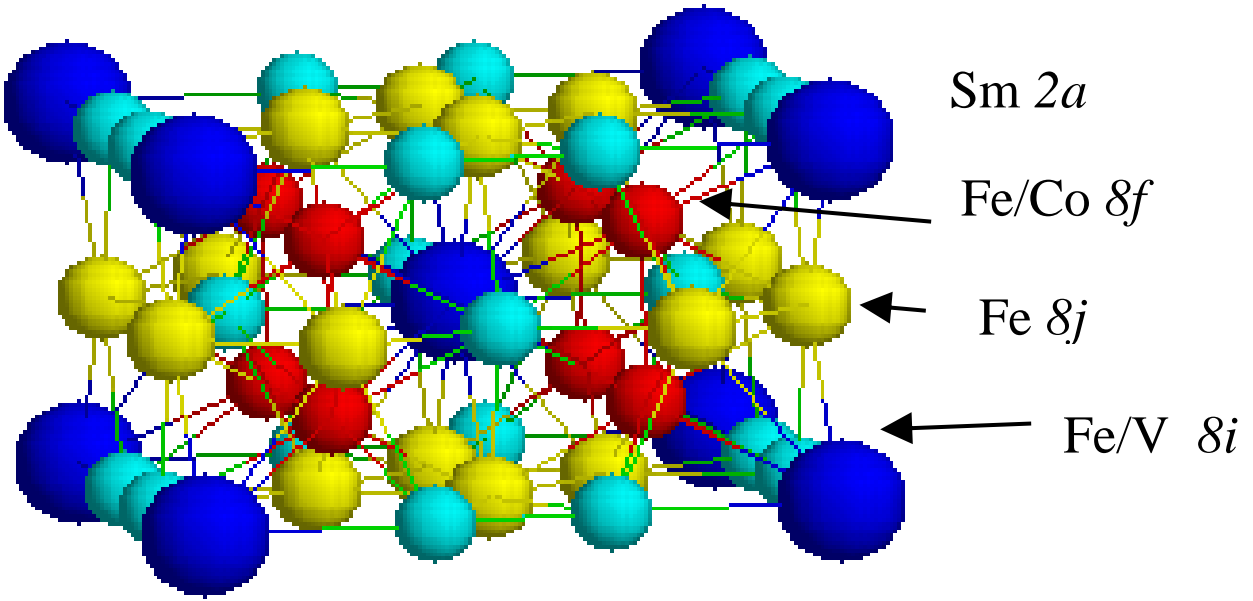

*Figure 1. The* $ThMn_{12}$ *structure (space group I4/mmm) of* $SmFe_{10}Co_{2-x}V_x$ *alloy. The atoms on 2a sites are dark blue spheres, whilst atoms on 8i, 8j, and 8f sites are represented as light blue (8i), green (8j), and red (8f) spheres, respectively.*

3.1 Density of states

Band structure calculations by SPR-KKR band structure method have been performed for $SmFe_{10}Co_{2-x}V_x$ (x = 0 - 2) alloys. The density of states (DOS) calculated for $SmFe_{10}V_2$ alloy by LSDA and LSDA+U approach, respectively, are presented in Fig. 2. The DOS plots obtained for $SmFe_{10}Co_2$ and $SmFe_{10}VCo$ alloys are available in the supplementary material. As can be seen in Fig. 2, the LSDA fails to describe the strongly localized *4f* orbitals of Sm, which are pinned at the Fermi level, set as the origin of the energy scale. By the LSDA+U approach, the *f*-orbitals of Sm are split into lower and upper Hubbard bands. The Sm-*4f* bands are not occupied in the spin-up channel, being occupied only in the spin-down channel, confirming that the Sm spin moments align opposite to the Fe spin moments. The occupied and unoccupied *4f*-bands of Sm are separated by $U$ - $J$ energy value. By LSDA calculations, no significant changes in DOS of transition metals can be observed. Still, the Sm-*4f* spin-up bands are above the Fermi level, resulting in a similar coupling between Sm and Fe spins.

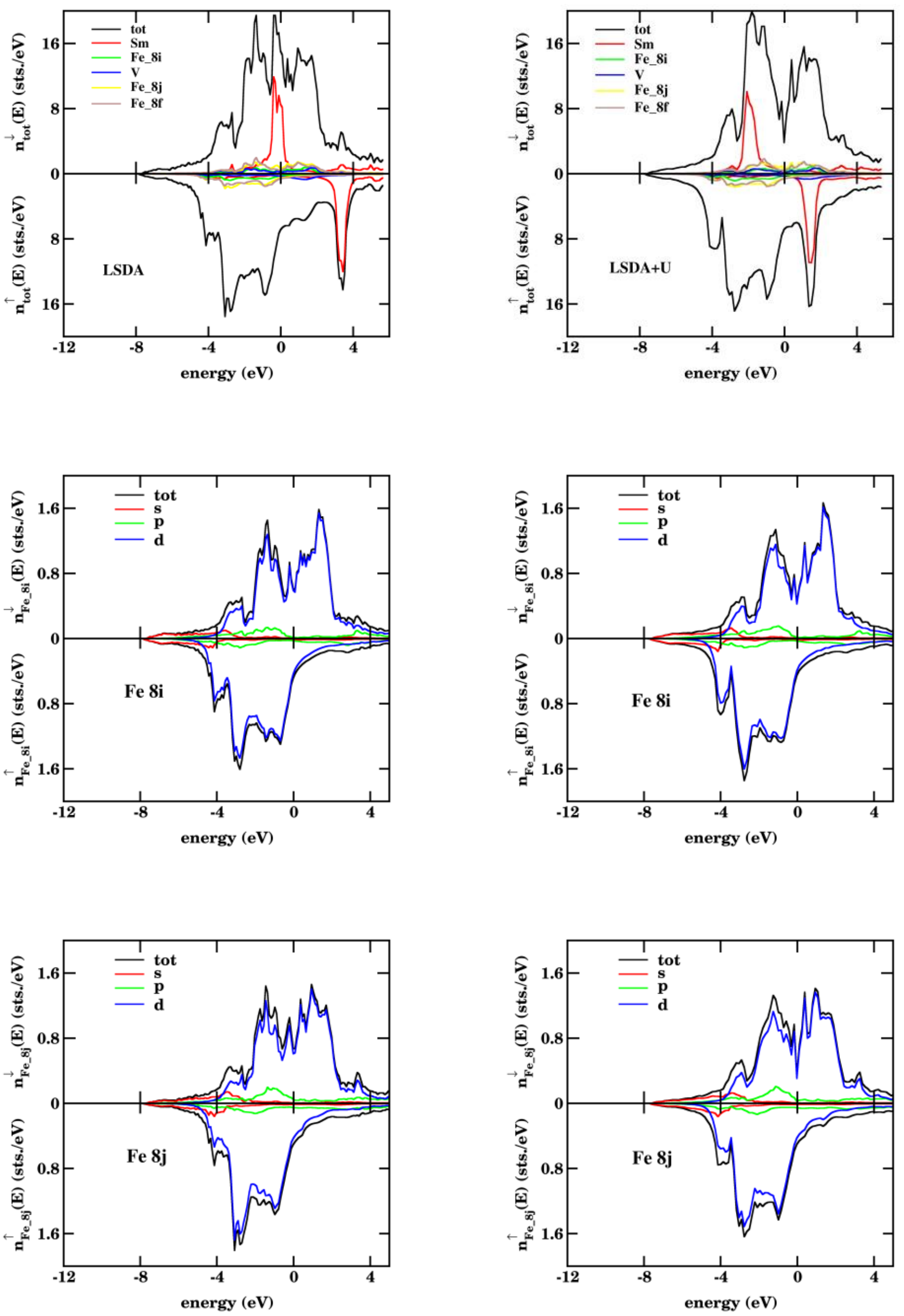

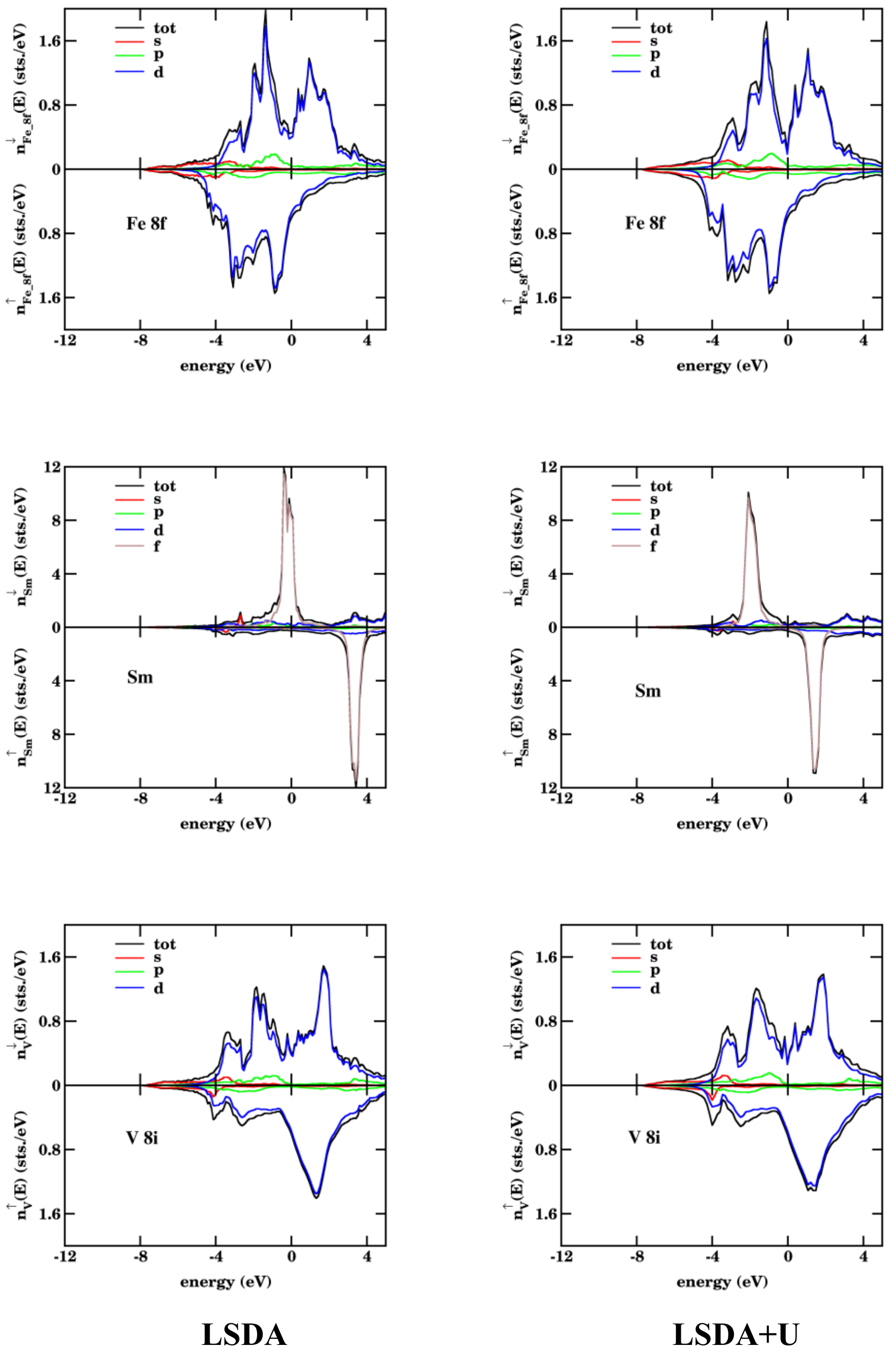

*Figure 2. The density of states of $SmFe_{10}V_2$ alloy by LSDA (left) and LSDA+U (right) approach, respectively. The Fermi level is the origin of the energy scale.*

### 3.2 Magnetic moments

The calculated magnetic moments for $SmFe_{10}Co_{2-x}V_x$ (x = 0 - 2) alloys are shown in Table 2. The Sm spin moment absolute values obtained by the LSDA approach for the investigated alloys are close to the spin magnetic moment in the ideal case for $4f^6$ electronic configuration, which is 6 $\mu_B$. The Sm spin moments range between 5.66 $\mu_B$ for $SmFe_{10}Co_2$ and 5.72 $\mu_B$ for $SmFe_{10}V_2$ alloy, respectively. The Sm spin moments absolute values are smaller by the LSDA+U approach (5.15 $\mu_B$ for $SmFe_{10}V_2$ and 5.18 $\mu_B$ for $SmFe_{10}CoV$ and $SmFe_{10}Co_2$ alloys). The Sm spin moment by the LSDA+U approach is close to the value obtained by Odkhuu et al. for $SmFe_{10}Co_2$ alloy (-5.13 $\mu_B$) [23] by using similar LSDA+U approach and the optimized lattice constants. The Sm-*4f* spins are coupled with the Sm-*5d* spins by intraatomic exchange interactions, their spins being oriented parallel to each other. The Sm-*5d* orbitals are more spatially extended, enabling hybridization with the Fe/Co-*3d* orbitals and their antiparallel spin coupling. As a consequence, the Fe/Co *3d* spins are antiparallel coupled to the Sm-*4f* spins [32], as suggested by the DOS plots (Fig. 2).

*Table 2. Calculated magnetic moments (in Bohr magnetons $\mu_B$) for $SmFe_{10}V_2$, $SmFe_{10}Co_2$ and $SmFe_{10}VCo$ alloys. Preferential occupation of Co 8f sites and V for 8i, respectively, has been considered.*

| | $SmFe_{10}V_2$ | | | | $SmFe_{10}CoV$ | | | | $SmFe_{10}Co_2$ | | | |
|---|---|---|---|---|---|---|---|---|---|---|---|---|
| | LSDA | | LSDA+U | | LSDA | | LSDA+U | | LSDA | | LSDA+U | |
| | $m_s(\mu_B)$ | $m_l(\mu_B)$ | $m_s(\mu_B)$ | $m_l(\mu_B)$ | $m_s(\mu_B)$ | $m_l(\mu_B)$ | $m_s(\mu_B)$ | $m_l(\mu_B)$ | $m_s(\mu_B)$ | $m_l(\mu_B)$ | $m_s(\mu_B)$ | $m_l(\mu_B)$ |
| Sm | -5.72 | 2.98 | -5.15 | 4.78 | -5.67 | 2.78 | -5.18 | 4.79 | -5.66 | 2.66 | -5.18 | 4.79 |
| Fe *8i* | 2.03 | 0.07 | 2.20 | 0.06 | 2.28 | 0.07 | 2.27 | 0.06 | 2.39 | 0.07 | 2.37 | 0.05 |
| V *8i* | -0.89 | 0.03 | -0.99 | 0.02 | -1.24 | 0.03 | -1.25 | 0.02 | - | - | - | - |
| Fe *8j* | 2.08 | 0.08 | 2.05 | 0.06 | 2.22 | 0.08 | 2.23 | 0.06 | 2.45 | 0.08 | 2.42 | 0.06 |
| Fe *8f* | 1.60 | 0.05 | 1.75 | 0.04 | 1.83 | 0.06 | 1.89 | 0.04 | 2.04 | 0.05 | 2.05 | 0.03 |
| Co *8f* | | | | | 1.41 | 0.09 | 1.45 | 0.08 | 1.57 | 0.09 | 1.59 | 0.08 |
| Total/f.u. | 11.30 | 3.71 | 12.44 | 5.32 | 15.71 | 3.61 | 16.42 | 5.43 | 20.91 | 3.53 | 21.25 | 5.47 |

On the other hand, the Sm orbital moment is antiparallel with the Sm spin moment, according to Hund's third rule for less than half-filled 4f-shells [32,33]. The LSDA+U values for the Sm orbital moments in the $SmFe_{10}Co_{2-x}V_x$ (x = 0-2) alloys are between 4.78 - 4.79 $\mu_B$, being almost independent of the V and Co samples content. The LSDA+U calculated values compare well with the Sm orbital moment obtained by Hund's rule (5 $\mu_B$). The LSDA calculations underestimate the orbital moments of Sm. The LSDA calculated values of orbital moments range between 2.66 $\mu_B$ for $SmFe_{10}Co_2$ and 2.98 $\mu_B$ for $SmFe_{10}V_2$. The orbital moment values obtained by our LSDA+U calculations are comparable with the Sm-*4f* orbital moments obtained by Liu et al. in $Sm(Fe,Co)_5$ (4.89 $\mu_B$) [34].
Other calculations using the FLAPW GGA+U method for the $SmFe_{10}Co_2$ alloy obtained a lower value of the orbital moment (2.25 $\mu_B$) [23]. Our investigations found for the resultant total magnetic moment of Sm atom by LSDA+U calculations values between 0.37 and 0.39 µB, slightly lower than the theoretical values of the $Sm^{3+}$ ion derived from the Hund's third rule (0.71 µB). Considering the notable difficulties associated with rare earth elements description by first-principles methods, this result is reasonable.

The Fe spin magnetic moments have different values on the three crystallographic sites, with the lowest value of the $m_s^{Fe}$ (*8f*) being related to the shortest interatomic distances around the *8f* site (Table 1). The largest value of the spin moment is $m_s^{Fe}$ (*8i*) for $SmFe_{10}V_2$ and $SmFe_{10}CoV$ samples; a similar magnitude sequence is found for $YFe_{11}Ti$ alloys by Mössbauer measurements [31]. Interatomic distances rise in $SmFe_{10}Co_{2-x}V_x$ alloys when V substitutes for Co due to an increase in the lattice constant $a_{lat}$. As can be seen in Table 1, the Fe *8j* and *8i* muffin-tin radii are the most affected by the lattice constant increase, while the Fe 8f muffin-tin radius (and implicitly the shortest interatomic distance) is almost unchanged. The overlap of the Fe/Co *3d* states is decreasing due to lattice growth, which is going to decrease exchange interactions. Consequently, for every Fe crystal site in the samples, the spin magnetic moments decrease by V for Co substitution.

*Table 3. The experimental and calculated $\mu_0 M_s$ (T) for investigated alloys.*

| | $SmFe_{10}V_2$ | $SmFe_{10}VCo$ | $SmFe_{10}Co_2$ |
|---|---|---|---|
| $\mu_0 M_{s\,exp}(T)$ | 0.81[a] [10]<br>0.9[a, b] [11] | - | 1.78[a, c] [6]<br>1.88[d] |
| $\mu_0 M_{s\,theor}^{LSDA+U}(T)$ | 1.19<br>0.93 [24] | 1.48 | 1.84<br>1.71[e] [23]<br>1.68[e] [24] |
| $\mu_0 M_{s\,theor}^{LSDA}(T)$ | 1.01 | 1.32 | 1.68 |

[a] at 300 K
[b] $SmFe_{10.1}V_{1.9}$ alloy at 300K
[c] the composition of the alloy was $SmFe_{9.6}Co_{2.4}$
[d] extrapolated at 0K
[e] by FLAPW method using GGA+U approach; U = 6 eV (Wien2K calculation code) considering the preferential occupation of Co for *8f* sites

Moreover, the Hubbard U correction raises the Fe/Co spin magnetic moment values for every alloy under investigation. Comparable values for Fe spin moments (2.58 $\mu_B$ on *8i*, 2.50 $\mu_B$ on *8j,* and 2.18 $\mu_B$ on *8f*) have been obtained by Ochirkhuyag et al. by the FLAPW GGA+U method for the $SmFe_{10}Co_2$ alloy [24]. When comparing the results of the LSDA(+U) calculations, the known GGA(+U) magnetic moment enhancement must be taken into account.

The magnitude of the Co spin moments increases with Co addition, being comparable with the values found by Odkhuu et al. [23] in $SmFe_{10}Co_2$ (1.38 $\mu_B$). The orbital moments of all *3d* atoms calculated by LSDA are reduced by applying the Hubbard U correction. Important orbital magnetic moment values have been obtained for Co *8f* (0.08-0.09 $\mu_B$) for both LSDA and LSDA+U approaches.

Experimental studies [6,10,11] reveal that the V for Co substitution decreases the magnetization in $SmFe_{10}Co_{2-x}V_x$ (x = 0-2) alloys, which is consistent with the current theoretical results. Using the calculated magnetic moments, the saturation magnetization $\mu_0M_s$ (T) values have been derived and compared with experimental measurements and other available theoretical calculations in Table 3. The calculated $\mu_0M_s$ values show an expected overestimation in comparison with the experimental results, as the magnetization measurements have been done at 300 K [ 6,10,11]. The agreement between theoretical and experimental data improves considerably when the 0K extrapolated magnetization for the $SmFe_{10}Co_2$ alloy is considered [6]. The $\mu_0 M_{s\,theor}^{LSDA+U}$ saturation magnetization values are significantly enhanced, improving the agreement with experiment when compared to prior theoretical results [23,24]. This is due to the larger orbital moment of Sm included in the total magnetic moment via LSDA+U method. The magnetization values obtained for $SmFe_{10}Co_2$ alloy are higher than the magnetization of the $Nd_2Fe_{14}B$ alloy, considered the best commercial magnet (1.61 T) [35]. A relatively high value of magnetization ($\mu_0M_s$ = 1.48 T) is predicted for the $SmFe_{10}VCo$ alloy.

### 3.3 Magnetocrystalline anisotropies

The magnetocrystalline energy (MAE) is an essential quantity in permanent magnets, representing the required energy to change the magnetization direction from the easy axis to the hard axis of magnetization by applying a magnetic field. In the $Sm(Fe,M)_{12}$ magnetic materials, the MAE originates from the Fe/M sub-lattice and the Sm-*4f* orbital contribution due to the spin-orbit coupling and crystal field effects [1]. It is well accepted that in these compounds the anisotropy is largely single-ion, and each rare earth and transition metal element yields an individual anisotropy contribution, with the alloy MAE determined as the sum of the transition metal sublattice and the Sm-sublattice anisotropy.

To investigate the microscopic origin of MAE, the magnetic anisotropy constants $K_u$ calculated by the magnetic torque method for the $SmFe_{10}Co_{2-x}V_x$ (x = 0 - 2) alloys have been decomposed into the Sm contribution (sitting on the *2a* crystallographic sites) and the contributions of Fe/M atoms (M = Co, V) sitting on the *8i*, *8j* and *8f* crystallographic sites (Fig. 3).

In crystal field theory, the rare earth anisotropy constant can be approximated in the lowest order by

$$K_u = -3J(J-1)\alpha_J \langle r^2 \rangle A_2^0 n_R \qquad (3)$$

where J is the total angular momentum, $\alpha_J$ is the first Stevens coefficient, $\langle r^2 \rangle$ is the spatial extent of the *4f* orbital, $A_2^0$ is the second-order crystal-field parameter, and $n_R$ is the rare-earth concentration [32]. The spin-orbit coupling of rare earth is strong enough to rigidly couple the rare earth spin to the aspherical charge distribution of the *4f* shell. The charge distribution of the Sm *4f* orbitals has a prolate shape with respect to the c-axis, corresponding to a positive second-order Stevens factor [32]. The charge distribution is subject to the action of the crystal field created by the non-spherical local environment of surrounding ions. The crystal field is characterized by the second-order crystal-field parameter $A_2^0$, which is positive if the charge along the c-axis is negative (repulsive for the electrons) and negative otherwise [32]. For all investigated $SmFe_{10}Co_{2-x}V_x$ (x = 0 - 2) alloys, the second-order crystal-field parameter $A_2^0$ is negative, resulting in an axial anisotropy contribution from Sm atoms.

As can be seen in Fig. 3(a), the Sm has the main contribution to the MAE for all the investigated systems, using both LSDA and LSDA+U calculation approaches. Within the LSDA approach, the Sm 4f states are described as fully itinerant valence states. A partial localization of the 4f stated can be expressed using the LSDA+U. Still, the LSDA+U approach must be a more realistic method as it is placing the *4f* states closer to where they belong (away from the Fermi level). The LSDA+U calculations found that $K_u$ of Sm increased by Co for V substitution, in agreement with the findings of Yoshioka et al. [36]. The substituted Co attracts electrons, which reduces the electron distribution in the interstitial region. The Sm *4f* charge distribution is strongly fixed along the c-axis, which enhances the MAE, as can be seen in Fig. 3(a).

For the *3d* contribution to the MAE, there are not well-defined rules, as the $K_u$ can oscillate as a function of the band filling [37]. Fig. 3 (b-d) shows the 3d elements contribution MAE, considering the Fe/V atoms on crystallographic site 8i (b), Fe atoms on crystallographic site 8j (c), Fe/Co atoms on crystallographic site 8f (d). Our calculations show that the $K_u$ contributions of Fe/M atoms (M = Co, V) are an order of magnitude smaller than the $K_u$ contributions of Sm, with the maximum values on 8j crystal sites (Fig. 3(c)). We notice that LSDA+U calculations enhance the Fe/M atom contributions compared with the LSDA, more pronounced for the *8i* and *8j* sites. Also, the Fe/Co contributions of *8f* sites are negative for all the investigated alloys (Fig.3d) by both LSDA/LSDA+U calculation methods, in agreement with Odkhuu et al. [23].

Within the LSDA+U approach, our site-resolved calculated $K_u$ values for $SmCo_{10}Co_2$ alloy are in qualitative agreement with the calculations of Odkhuu et al. based on FLAPW GGA+U calculations

[23] (-0.09/-0.16 meV for Fe/Co *8f*, 0.04 meV for Fe *8i*, 0.10 meV for Fe *8j* and 20.5 meV for Sm). The calculation approach details explain the differences observed in the MAE calculations. In particular, the use of atomic sphere approximation (ASA) in our calculations neglects the anisotropic crystal field effects, which arise from the non-spherical potential, accounted for by full potential approach [23]. On the other hand, the $K_u$ values calculated within the general gradient approximation (GGA) often gives underestimated values with respect to the results obtained by local density approximation (LSDA) in other $RFe_{11}Ti$ (R=Y, Ce, Nd) alloys [38-40].

The calculated $K_u$ values have been compared with the experimental measurements [8,11,23,30] in Table 4. One has to note that by torque calculation method, the $E_{[100]} - E_{[001]}$ energy difference is determined [26], corresponding to the calculated anisotropy constant $K_u \approx K_1 + K_2$ determined by the experiment. Temperature should be considered when comparing the experimental anisotropy constants, since the rare earth sublattice contribution dominates at low temperatures while the transition metal sublattice contribution prevails at high temperatures. Only for $SmFe_{9.6}Co_{2.4}$ alloy, the comparison with the experimental measurement is straightforward, as the measurements are performed at 10 K and the theoretical prediction is for zero absolute temperature. The value of $K_u^{LSDA+U}$ (10.61 MJ/m$^3$) for the $SmFe_{10}Co_2$ alloy is in very good agreement with the value from the experimental measurements by Ogawa et al. at 10 K (9.73 MJ/m$^3$) [8] and with the FLAPW GGA+U calculated value (11.2 MJ/m$^3$) [23].

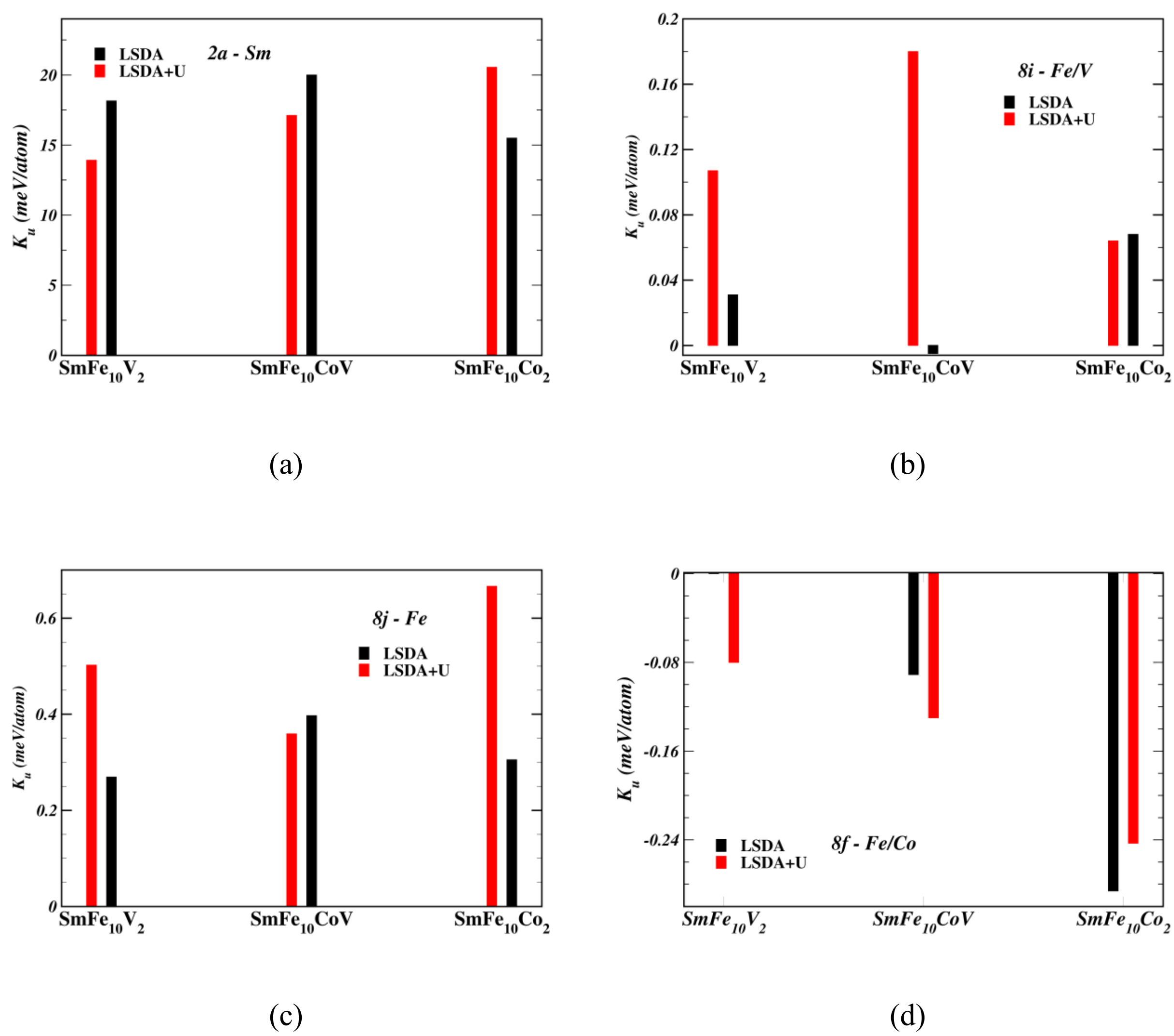

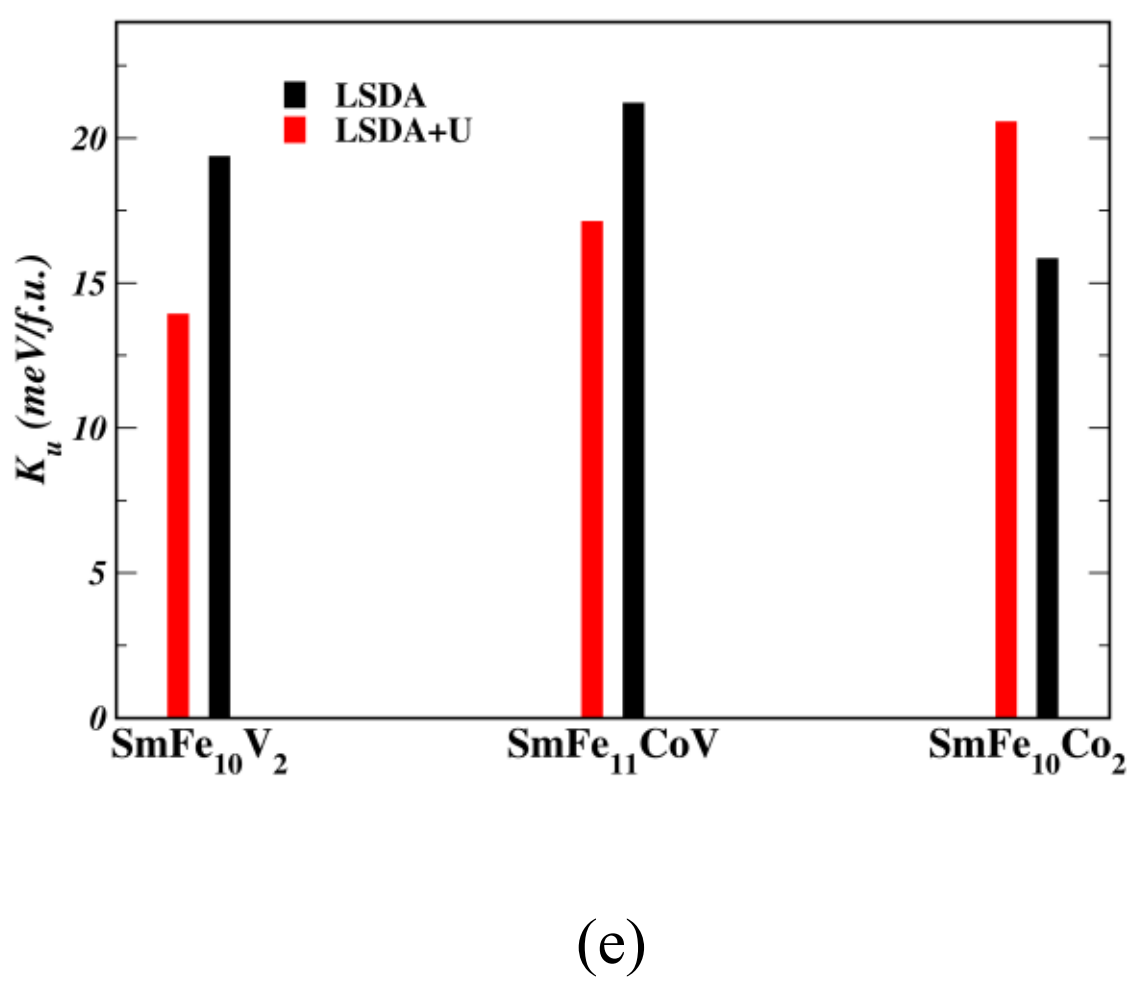

(e)

*Figure 3. The magnetic anisotropy constants $K_u$ calculated by the magnetic torque method by LSDA/LSDA+U method for the* $SmFe_{10}Co_{2-x}V_x$ *(x =0 - 2) alloys, decomposed into Sm -4f contributions (a), Fe/V atoms on crystallographic site 8i (b), Fe atoms on crystallographic site 8j (c), Fe/Co atoms on crystallographic site 8f (d). The total anisotropy constants $K_u$ (in meV/f.u) are shown in (e).*

*Table 4. The experimental and calculated anisotropy constants, anisotropy field $H_a$ and magnetic hardness κ for investigated alloys.*

| | $SmFe_{10}V_2$ | $SmFe_{10}VCo$ | $SmFe_{10}Co_2$ |
|---|---|---|---|
| $K_u^{exp}$ (MJ/m$^3$) | 5.22 [11] [a]<br>6.24 [30] [b] | | 9.73 [8] [c] |
| $K_1^{exp}$ (MJ/m$^3$) | 4.9 [11] [a] | | 11.86 [8] [c] |
| $K_2^{exp}$ (MJ/m$^3$) | 0.32 [11] [a] | | -2.13 [8] [c] |
| $K_u^{LSDA+U}$ (MJ/m$^3$) | 7.53 | 9.12 | 10.61<br>11.2 [23] |
| $K_u^{LSDA}$ (MJ/m$^3$) | 8.93 | 9.91 | 15.83 |
| $\mu_0 H_a^{LSDA+U} (T)$ | 15.6 | 15.5 | 14.5 |
| $\kappa^{LSDA+U}$ | 2.57 | 2.28 | 1.34 |

[a] determined at 100 K

[b] determined at RT
[c] determined at 10 K for $SmFe_{9.6}Co_{2.4}$ alloy

The theoretical calculations found that the anisotropy constants decreased by V for Co substitution, but experimental confirmation of this finding is not possible due to measurement temperature for $SmFe_{10}V_2$ alloy. We note that the LSDA method tends to overestimate the anisotropy constant values, which are substantially adjusted by the LSDA+U method, improving the comparison with experimental data. The description of the Sm 4f electronic state by LSDA+U is essential for MAE calculations in $SmFe_{10}Co_{2-x}V_x$ (x = 0-2) alloys, in accordance with the results of Herper et al. [41] for magnetic anisotropy studies in $NdFe_{11}Ti$ alloy.

The anisotropy field was calculated based on the formula $H_a = 2K_u/\mu_0 M_s$ [37], with the values determined for uniaxial anisotropy constant $K_u$ (Table 4) and the magnetization $\mu_0 M_s$ (Table 3). The anisotropy field values indicate the theoretical maximum that coercivity can achieve, which requires a proper microstructure. Given the empirical Kronmüller equation's link between anisotropy and coercivity [42], the larger anisotropy fields found for each of these alloys assume the fabrication of magnets with higher coercivity. Since the specific magnet's microstructure, including pinning and nucleation centers, is crucial the reported experimental measurements found coercivity $H_c$ values of only 10% of the $H_a$ value in $Sm(Fe,M)_{12}$ alloys [43]. However, $Sm(Fe,M)_{12}$ alloys with M = V have been found to have relatively high coercivity experimentally [14], which is in line with our theoretical estimates that indicate greater anisotropy fields for V-containing alloys when compared to $SmFe_{10}Co_2$ alloy. Moreover, the present theoretical results show for the anisotropy field of the hypothetical $SmFe_{10}VCo$ alloy a relatively high value (15.5 T). Also, the calculated anisotropy constant of the $SmFe_{10}VCo$ alloy (9.12 $MJ/m^3$) is exceeding that of other predicted alloys, such as $SmFe_{10}CoAl$ (7.8 $MJ/m^3$) and $SmFe_9Co_2Al$ (8 $MJ/m^3$) [24], respectively.

Using the LSDA+U calculated values of $\mu_0 M_s$ and the anisotropy constants $K_u^{theor} \approx K_1 + K_2$, magnetic hardness parameters of the alloys $\kappa = \sqrt{K_u^{theor}/\mu_0 M_s^2}$ have been calculated [37] (Table 4). Skomski et al. [37] state that having a hardness parameter $\kappa > 1$ is a general rule of thumb for a bulk material, regardless of shape, to resist self-demagnetization and to be used as permanent magnet. The $\kappa^{LSDA+U}$ calculated values range between 2.57 for $SmFe_{10}V_2$ and 1.34 for $SmFe_{10}Co_2$. The calculated hardness parameter for the $SmFe_{10}Co_2$ alloy is comparable with the value determined by Ogawa et al. for epitaxially grown thin films ($\kappa = 1.24$) [8]. The hardness parameters for the $Sm(Fe,M)_{12}$ alloys are enhanced by the increase of the V:Co ratio in the investigated $SmFe_{10}Co_{2-x}V_x$ (x = 0-2) alloys. As a consequence, the $SmFe_{10}Co_{2-x}V_x$ alloys with higher V content are more suitable to build hard permanent magnets.

### 3.4 Curie temperatures

The Curie temperatures deduced by mean-field approximation (MFA) [27-29] are presented in Table 5. The Curie temperature is mainly determined by the Fe - Fe exchange interactions, inducing a ferromagnetic coupling for all investigated alloys [6,10,11]. For an increased content of V, the number of Fe next neighbours on each Fe site is lowered, leading to a weaker Fe - Fe exchange. This is seen in the lower *Tc* values for $SmFe_{10}V_2$ alloy (680 K) compared to $SmFe_{10}CoV$ (910 K) alloy.

*Table 5. Curie temperatures were estimated by mean-field approach [27-29] for $SmFe_{10}V_2$, $SmFe_{10}Co_2$, and $SmFe_{10}VCo$ alloys, together with experimental values [1,6,10,11].*

| | $SmFe_{10}V_2$ | $SmFe_{10}VCo$ | $SmFe_{10}Co_2$ |
|---|---|---|---|
| $T_c^{exp}$(K) | 594 [10]<br>600 [11] [a]<br>610 [1] | - | 859 [6] [b] |
| $T_c^{theor}$ (K) | 680 | 910 | 1120 |

[a] for $SmFe_{10.1}V_{1.9}$ alloy

[b] for $SmFe_{9.6}Co_{2.4}$ alloy

On the other hand, the shrinkage of the lattice by Co for V substitution is causing an increase in the overlap of *3d* states of Fe/Co which would increase the exchange interactions [43]. This is reflected by the high value of the $T_c$ for $SmFe_{10}Co_2$ (1120 K) compared with the $SmFe_{10}V_2$ alloy.

The present MFA approach only allows for a qualitative evaluation of the Tc dependency on the V:Co ratio because the MFA overstates the Curie temperature (Tc) by around 20% [44]. Indeed, by our MFA calculations the $T_c$ value for the $SmFe_{10}Co_2$ alloy was overstated (1120K) when compared to the experimental value of 859 K. The main reasons for these MFA overestimated values are known, and they are linked to (i) unaccounted long-range interactions and (ii) the exchange-correlations parameters $J_{ij}$ used to compute the Tc obtained for zero absolute temperature [44]. More reliable estimations based on the Monte-Carlo approach or random phase approximation (RPA) should give closer theoretical values for the Curie temperatures.

The value $T_c^{theor}$ predicted for the $SmFe_{10}VCo$ alloy is 910 K, considerably higher than that of $SmFe_{10}V_2$ alloy. Accounting for the MFA overestimation of $T_c^{theor}$, this alloy is expected to have a Curie temperature over 730 K, making it suitable for permanent magnet applications.

## 4. Conclusions

The present studies show the intrinsic magnetic properties of the $SmFe_{10}Co_{2-x}V_x$ (x = 0-2) derived from the SPR-KKR LSDA(+U) theoretical calculations. The theoretical calculated values of magnetization and magnetic anisotropy energy, particularly those generated using the LSDA+U method for the $SmFe_{10}Co_2$ alloy, are in agreement with earlier theoretical results [23,24] as well as experimental magnetization and anisotropy studies at low temperatures [6,8]. Considering the known overestimation of the Curie temperatures by MFA, the calculated $T_C^{theor}$ values show reasonable agreement with experimental values for both $SmFe_{10}Co_2$ [6] and $SmFe_{10}V_2$ alloys [1,10,11]. The hypothetical $SmFe_{10}CoV$ alloy retained a high magnetization, anisotropy, and Curie temperature while having a greater magnetic hardness value than the $SmFe_{10}Co_2$ alloy [8]. As a result, the shortcoming found for $SmFe_{10}Co_2$ alloys, which can be obtained only as epitaxial films, requiring in addition special methods to develop increased coercivity [8, 9], can be solved by limited V for Co doping (x = 1). Previous research have obtained a similar $Sm(Fe,V)_{12}$ alloy with low V content[10]. The as-obtained alloy shows potential to develop a stable hard magnetic phase, and it is suggested for further experimental exploration.

## Acknowledgments

DB acknowledges Ondrej Sipr for the fruitful discussions as well as for technical support related to the LSDA+U SPR-KKR approach.

**Funding:** This publication was supported by the project Quantum materials for applications in sustainable technologies (QM4ST), funded as project No. CZ.02.01.01/00/22_008/0004572 by Programme Johannes Amos Commenius, call Excellent Research. Fundings from the Ministry of Research, Innovation and Digitization by CCCDI-UEFISCDI grant PN-III-P2-2.1-PED-2019-3484 are also acknowledged.

**Data Availability Statement:** Data are available on request.

***Supplementary material***

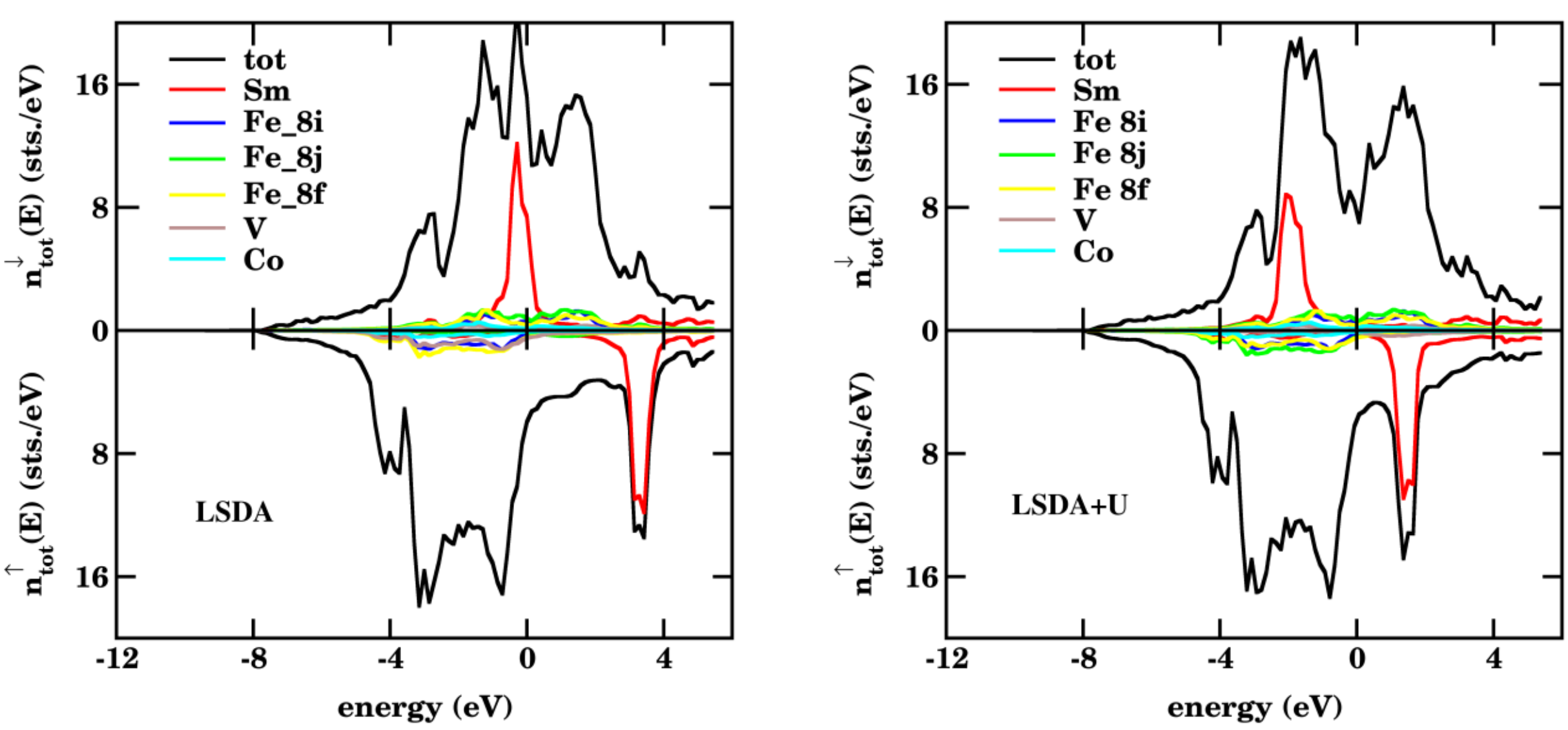

Fig. A. Density of states calculated by LSDA and LSDA+U approach, respectively, for the $SmFe_{10}CoV$ alloy.

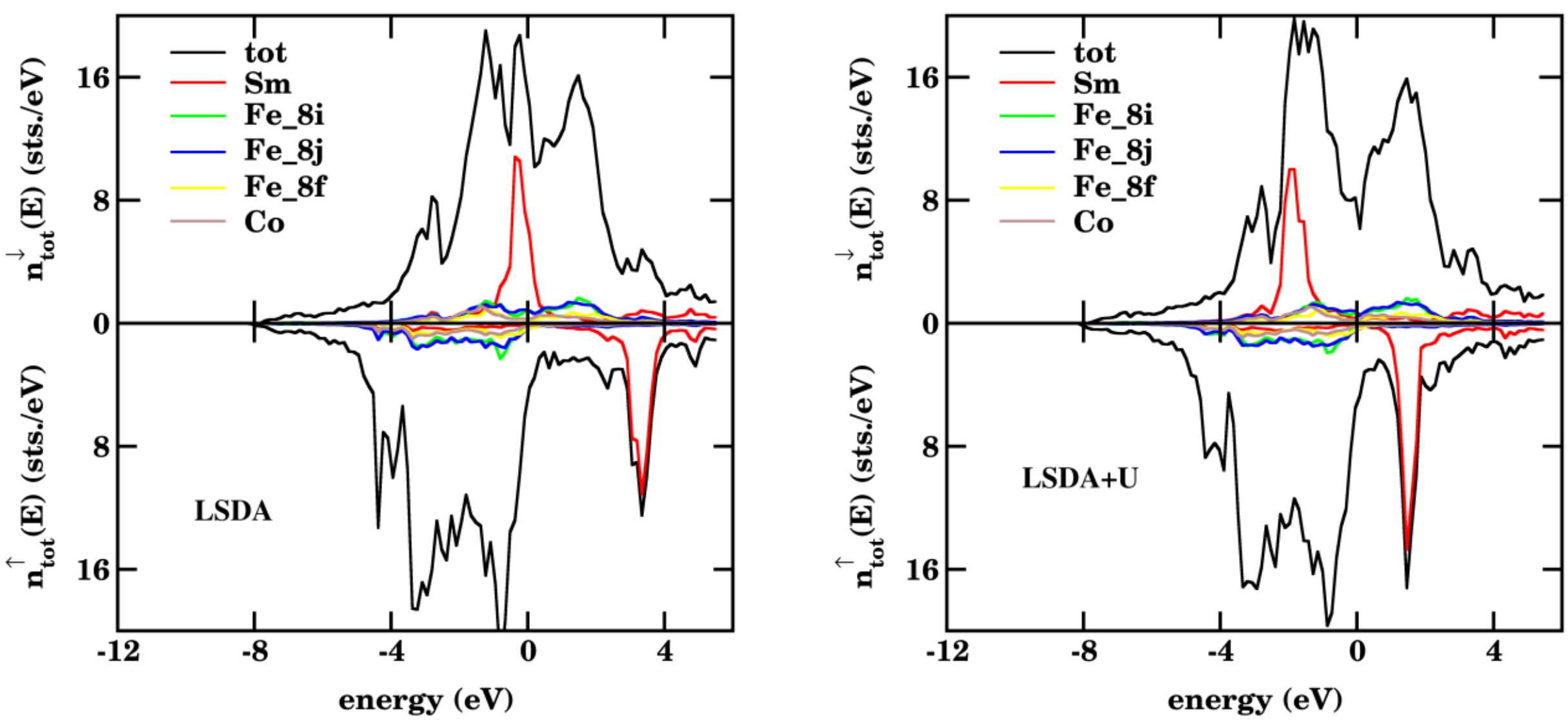

Fig. B. Density of states calculated by LSDA and LSDA+U approach, respectively, for the $SmFe_{10}Co_2$ alloy.